\documentclass[reprint, amsmath, amssymb, superscriptaddress]{revtex4-2}

\usepackage{graphicx}% Include figure files
\usepackage[font={small, it}]{caption}
\usepackage{float}
\usepackage{natbib}
\usepackage{amsmath}
\usepackage{dcolumn}% Align table columns on decimal point
\usepackage{bm}% bold math
\usepackage[colorlinks=true,citecolor=blue,linkcolor=blue,urlcolor=blue]
{hyperref}
\begin{document}

	\title{Does violation of Cosmic No-hair Conjecture guarantee the existence of Wormhole?}

\author{Dhritimalya Roy}
\email{rdhritimalya@gmail.com}
\affiliation{Department of Physics, Jadavpur University, Kolkata-700032, INDIA}

\author{Ayanendu Dutta}
\email{ayanendudutta@gmail.com}
\affiliation{Department of Physics, Jadavpur University, Kolkata-700032, INDIA}

\author{Subenoy Chakraborty}
\email{schakraborty.math@gmail.com}
\affiliation{Department of Mathematics, Jadavpur University, Kolkata-700032, INDIA}

	\begin{abstract}
	The present work investigates the interrelation between the validity (or violation) of the cosmic no-hair conjecture and the existence (or non-existence) of wormholes, both in Einstein's Gravity and in modified gravity theories. It is found that the existence of wormholes implies a violation of the cosmic no-hair conjecture, and the validity of the cosmic no-hair conjecture implies the non-existence of wormholes but not the reverse way both in Einstein's Gravity and in modified gravity theories. We will also re-visit the gravitational entropy predictions of the wormhole and show how they are connected.
\end{abstract}
\maketitle

\section{Introduction}\label{intro}
Wormholes are usually defined as smooth bridges between different universes or topological handles (i.e. throats having no horizons) between remote parts of a single universe. Earlier, wormholes are purely of theoretical curiosity \cite{morris1988, Lorentzian.Wormholes.Visser} but recently, this theoretical aspect has gained significant importance due to the present accelerating phase of our universe. There is a nice similarity between this theoretical phenomenon and the recent observational aspects. A traversable wormhole is supported by a so-called exotic matter which violates the null energy condition (i.e. $\rho+p<0$) at least in a neighborhood of the wormhole throat \cite{hochberg1998, hochberg1998_2}. Exotic matter or matter that violated the strong energy condition (i.e. $\rho+3p<0$) has started gaining importance in the last two decades because of the observed acceleration of the present-day universe. It is speculated that the recent acceleration in the expansion of the universe could only be explained using matter that is characterized by the negative equation of state, called dark energy (DE). A useful choice of DE is the phantom energy having equation of state $p=-\omega \rho$ ($\omega<-1$), thereby violating the null energy condition (NEC). Several papers on wormholes show that such phantom energy could be the nature of the matter of wormhole throat \cite{morris1988_2, lobo2005, lobo2005_2, cataldo2012, cataldo2013, folomeev2014}. Recently, numerous wormhole solutions are investigated in different modified gravity theories such as Bumblebee gravity, Teleparallel theories of gravity, etc. using various matter content. Gravitational lensing, light deflection angle, various thin-shell wormhole models have also been extensively studied. Also, several works have been carried out on the existence of a wormhole solution with non-exotic matter. Readers may refer to \cite{ali1,ali2,ali3,ali4,ali5,ali6,ali7,moraes,moraes2,matter1,matter2,matter3,matter4} for more details.

On the other hand, in general terminology, the Cosmic no-hair conjecture (CNHC) states that "all expanding universe models with a positive cosmological constant asymptotically approach the de-Sitter solution". To address the question of whether the universe evolves into a homogeneous and isotropic state during an inflationary epoch, Gibbons et al. \cite{gibbons1977} and Hawking et al. \cite{hawking1982} developed this conjecture. Wald \cite{wald1983} gave a formal proof of it using matter that is normal and satisfies the dominant energy condition (DEC) and strong energy condition (SEC). Later on, CNHC has been extensively used in several contexts and also in different gravity theories \cite{kitada1992, chakraborty2001, chakraborty2002, chakraborty2003, chakraborty2009}.

In this letter, the motivation of the present work is to study the CNHC as well as the formation of wormhole both in Einstein's gravity and in modified gravity theories. More specifically, we shall examine whether there is an interrelation between the formation of wormholes and the validity or violation of CNHC. For the formation of the wormhole, the matter at the throat should violate the null energy condition (NEC) (and hence DEC and SEC) while for the validity of CNHC, the matter must satisfy both the strong energy condition (SEC) and weak energy condition (WEC). So a natural question that comes to mind is: does violation of CNHC implies the formation of a wormhole or vice versa? We shall address this question both in Einstein's Gravity as well as in modified gravity theories. 

We studied the relation between CNHC and wormholes. Later, we take the Hayward formalism for describing the wormhole \cite{hayward1998, hayward2004, hayward2009} and show how by taking their method for describing the wormhole, past outer trapping horizon, we can verify the results.

\section{Energy Conditions: CNHC and Wormholes}\label{energy}
To analyze the cosmic evolution in Einstein's Gravity we consider the Hamiltonian constraint arising from the Einstein equation and the Raychaudhuri equation, which is given by,
\begin{equation}\label{eqn1}
	G_{\mu \nu} n^\mu n^\nu -\kappa^2 T_{\mu \nu} n^\mu n^\nu=0
\end{equation}
\begin{equation}\label{eqn2}
	R_{\mu \nu} n^\mu n^\nu -\kappa^2 (T_{\mu \nu} -\frac{1}{2} T g_{\mu \nu}) n^\mu n^\nu=0
\end{equation}
We now decompose the four dimensional manifold in (3+1) form so that the three space-like hypersurfaces are specified by the three metric $ h_{ab} $ given by,
\begin{equation}\label{eqn3}
	h_{ab} = g_{ab} + n_a n_b
\end{equation}
Here $ n^a $ be the unit normal vector to the hypersurface. The extrinsic curvature tensor $ K_{ab} = \nabla_a n_b  $ has the explicit form,
\begin{equation}\label{eqn4}
	K_{ab} = \frac{1}{3} K h_{ab} + \sigma_{ab}
\end{equation}
where $ K=K_{ab} h^{ab} $ is the trace of the extrinsic curvature and $ \sigma_{ab} $ is the shear of the time-like geodesic congruence orthogonal to the hypersurfaces. Now using the Gauss-Codazzi equation that relates the space-time intrinsic curvature to the three curvature of the hypersurface $ {}^3 R $, one obtains $ G_{\mu \nu} n^\mu n^\nu = \frac{1}{2}({}^3 R - K_{\mu \nu}K^{\mu \nu} + K^2) $. Rewriting the Ricci tensor in terms of the curvature tensor and then using some simple mathematical calculations, we obtain the dynamical equations (\ref{eqn1}) and (\ref{eqn2}) as
\begin{equation}\label{eqn5}
	K^2=3\kappa^2 T_{\mu \nu} n^\mu n^\nu + \frac{3}{2} \sigma_{\mu \nu} \sigma^{\mu \nu}-\frac{3}{2} {}^3 R
\end{equation}
\begin{equation}\label{eqn6}
	\dot{K}=-\frac{1}{3}K^2 - \sigma_{\mu \nu} \sigma^{\mu \nu} - \kappa^2 (T_{\mu \nu} -\frac{1}{2} T g_{\mu \nu}) n^\mu n^\nu
\end{equation}
where the overdot denotes the Lie derivative with respect to proper time.

It has been shown that the 3-space curvature scalar $ {}^3 R $ is negative definite for all Bianchi models except IX, for which $ {}^3 R $ is indefinite in sign. Thus using the idea of Wald and proceeding along the line of his approach one finds that for CNHC the matter should satisfy the weak and strong energy conditions, i.e.
\begin{equation}\label{eqn7}
	a\ge 0 ~~~~~and~~~~~b\ge 0
\end{equation}
where
\begin{equation}\label{eqn8}
	a= T_{\mu \nu} n^\mu n^\nu
\end{equation}
\begin{equation}\label{eqn9}
	b= (T_{\mu \nu} -\frac{1}{2} T g_{\mu \nu}) n^\mu n^\nu=a+ \frac{1}{2}T
\end{equation}

However, in modified gravity theories, the field equations can be written in the form of Einstein equations with an additional (hypothetical) matter term (known as geometric matter) as
\begin{equation}\label{eqn10}
	G_{\mu \nu}=T_{\mu \nu}+T^{(g)}_{\mu \nu}
\end{equation}
As a result, equations (\ref{eqn1}) and (\ref{eqn2}) take the form
\begin{equation}\label{eqn11}
	G_{\mu \nu}n^\mu n^\nu=T_{\mu \nu}n^\mu n^\nu+T^{(g)}_{\mu \nu}n^\mu n^\nu
\end{equation}
and
\begin{equation}\label{eqn12}
	R_{\mu \nu} n^\mu n^\nu = (T_{\mu \nu} -\frac{1}{2} T g_{\mu \nu}) n^\mu n^\nu + (T^{(g)}_{\mu \nu} -\frac{1}{2} T^{(g)} g_{\mu \nu}) n^\mu n^\nu
\end{equation}
Hence after (3+1)-decomposition as before the above dynamical equations become (modifications of equations (\ref{eqn5}) and (\ref{eqn6})) (with $ \kappa^2=1 $)
\begin{equation}\label{eqn13}
	K^2=-\frac{3}{2} {}^3 R + 3 \sigma^2 +3a + 3c
\end{equation}
and
\begin{equation}\label{eqn14}
	\dot{K}=-\frac{1}{3}K^2 -2 \sigma^2 -b-d
\end{equation}
where $ \sigma^2=\frac{1}{2} \sigma_{\mu \nu} \sigma^{\mu \nu} $ is the (square of) shear scalar, and the scalars $c$ and $d$ have the expressions
\begin{equation}\label{eqn15}
	c= T^{(g)}_{\mu \nu} n^\mu n^\nu
\end{equation}
\begin{equation}\label{eqn16}
	d= (T^{(g)}_{\mu \nu} -\frac{1}{2} T^{(g)} g_{\mu \nu}) n^\mu n^\nu=c+ \frac{1}{2}T^{(g)}
\end{equation}
Thus for the validity of the CNHC in modified gravity theories we must have
\begin{equation}\label{eqn17}
	a+c\ge 0 ~~~~~and~~~~~b+d\ge 0
\end{equation}
i.e. the effective matter (equivalent geometric matter + normal matter) must obey the weak and the strong energy conditions.

On the other hand, for wormhole configuration we consider the Raychaudhuri equation as
\begin{equation}\label{eqn18}
	\frac{d\theta}{d\tau}=-\frac{1}{3}\theta^2 -2 \sigma^2 +2\omega^2 - R_{\mu \nu} k^\mu k^\nu
\end{equation}
where $ \theta $ is the expansion scalar, $ \omega^2=\frac{1}{2} \omega_{\mu \nu}\omega^{\mu \nu} $ is the (square of) rotation scalar, and $ \tau $ is the proper time. Also here the shear and rotation scalars are associated to the congruence defined by the null vector field $ k^\mu $ (with $ k^\mu k_\mu=0 $). As for any hypersurface orthogonal congruences $ \omega_{\mu \nu}=0 $ so for attractive gravity, the above Raychaudhuri equation demands $ R_{\mu \nu} k^\mu k^\nu \ge 0 $. Hence the geodesic congruences focus within a finite value of the parameter labeling points on the geodesics. On the contrary, the role of wormhole throat is to defocus geodesic congruences (flaring-out condition). So for wormhole configuration it is essential to have $ R_{\mu \nu} k^\mu k^\nu < 0 $. However, the Raychaudhuri equation is a purely geometric statement, it is not related to any particular gravity theory. So the above flaring-out condition for the existence of a wormhole takes the form:

(i) Einstein's Gravity:
\begin{equation}\label{eqn19}
	a_n \equiv T_{\mu \nu} k^\mu k^\nu<0
\end{equation}
(violation of null energy condition)

(ii) Modified gravity theory:
\begin{equation}\label{eqn20}
	T_{\mu \nu} k^\mu k^\nu + T^{(g)}_{\mu \nu} k^\mu k^\nu<0~~~~~i.e.~~~ a_n+c_n<0
\end{equation}
Here $ a_n $ and $ c_n $ are the expressions for $ a $ and $ c $ for null geodesics. Thus in the modified gravity theory there are three possibilities for the existence of a wormhole, namely:

I) Both the usual matter and the equivalent geometric matter (EGM) violates null energy condition (NEC) and as a result, the resultant effective matter violates NEC (i.e. $ a_{n}<0 $ and $ c_{n}<0 $).

II) The usual matter violates NEC (i.e. $ a_{n}<0 $) but the EGM does not (i.e. $ c_{n}>0 $), still the inequality (\ref{eqn20}) holds, provided $ c_{n} $ has an upper bound $ |a_{n}| $ (i.e. $ c_{n}<|a_{n}| $).

III) The NEC is obeyed (i.e. $ a_{n} \ge 0 $) by the usual matter while it is violated by the EGM (i.e. $ c_{n}<0 $) and $ a_{n} $ has an upper bound $ |c_{n}| $ (i.e. $ a_{n}<|c_{n}| $).
\newline
Further, if we consider a congruence of time-like geodesics instead of null one then the above Raychaudhuri equation becomes
\begin{equation}\label{eqn21}
	\frac{d\theta_{T}}{d\tau}=-\frac{1}{3}\theta^2_{T} -2 \sigma^2_{T} +2\omega^2_{T} - R_{\mu \nu} v^\mu v^\nu
\end{equation}
where all the quantities with subscript $ T $ stand for the same quantities as above for the time-like geodesics and $ v^{\mu} $ is a unit time-like vector (i.e. $ v^{\mu}v_{\mu}=-1 $). Thus the defocusing condition for the time-like geodesic congruence is 
\begin{equation}\label{eqn22}
	R_{\mu \nu} v^\mu v^\nu < 0
\end{equation}
which takes the following form in gravity theories,

(i) Einstein's Gravity:
\begin{equation}\label{eqn23}
	(T_{\mu \nu} - \frac{1}{2}T g_{\mu \nu})v^\mu v^\nu<0~~~~~i.e.~~~ b<0
\end{equation}
(violation of strong energy condition)

(ii) Modified gravity theory:
\begin{equation}\label{eqn24}
	\begin{split}
		(T_{\mu \nu} - \frac{1}{2}T g_{\mu \nu})v^\mu v^\nu + (T^{(g)}_{\mu \nu} - \frac{1}{2}T^{(g)} g_{\mu \nu})v^\mu v^\nu < 0   
		\\ i.e.~~~b+d <0 ~~~~~ i.e.~~~ a+c+ \frac{1}{2}(T+T^{(g)})<0		
	\end{split}	
\end{equation}

As above, we have again three possibilities for the occurrence of the wormhole configuration in modified gravity theory.

A. Both the matter components violate NEC and the trace of the effective total matter energy-momentum tensor has an upper bound of $ 2|a+c| $ or equivalently both the matter components violate SEC.

B. The usual matter component violates SEC (i.e. $ b<0$) while  EGM obeys SEC (i.e. $d \ge 0$)  and\ we have $ 0 \le d < |b|$.

C. The SEC is satisfied by usual matter (i.e. $b \ge 0$) but not by the EGM (i.e. $d < 0$) and $b$ is confined to the interval ($0, |d|$).

Note that conditions III and C will be interesting in the present context. Condition III shows that in modified gravity theories it is possible to have a wormhole with normal matter but the null scalar $a_{n}$ is restricted in the range ($0, |c_{n}|$). Condition C shows that wormholes may be formulated with normal matter which satisfies the SEC in modified gravity theories but the scalar b should be restricted to ($0,|d|$). We shall now examine both Einstein's Gravity and modified gravity theories on how the validity/non-validity of CNHC is related to the existence/non-existence of wormhole configuration.

From the energy conditions, it is well known that:

i) Weak Energy Condition (WEC) implies null energy condition (NEC) but not the converse.

ii) SEC does not imply weak energy condition (WEC) and vice versa.
\newline
Thus in Einstein's Gravity if the space-time satisfies CNHC then both WEC and SEC are satisfied and hence NEC is satisfied. So it is not possible to form wormholes in space-time where CNHC is satisfied. On the other hand, if wormholes form in space-time then matter should be exotic (i.e violates NEC). As a result, WEC will not be satisfied and hence CNHC will be violated. Therefore, in Einstein's Gravity we can say that: 
\begin{itemize}
	\item validity of CNHC  $ {\Rightarrow} $ non-existence of wormhole
	\item existence of Wormhole $ {\Rightarrow} $ violation of CNHC,
\end{itemize} 
but not in a reverse way, i.e non-existence of wormholes only implies that NEC is satisfied which does not mean that WEC  is satisfied. Similarly, a violation of CNHC may be due to a violation of SEC which does not imply a violation of NEC.

It should be noted that similar results also hold in modified gravity theories but with one basic difference. In Einstein's Gravity, the existence of a wormhole implies a violation of CNHC and hence the matter field must be exotic. On the other hand, in modified gravity theories we have the same conclusion but the matter field may not have an exotic character.

However for the wormhole, to make it traversable there are restrictions on the geometry which defines the throat. The wormhole throat is obtained by reconditioning two asymptotically flat space-times, such that the location of the throat is at their minimum value, and as one moves away from the throat there is a flare out of the space-time that has to be satisfied by the radial coordinate. This essentially means that congruence of geodesics that one obtains moving towards the focal point of an asymptotically flat metric is never achieved on either side of the restructured point (in this case the point of minimum $ r $ is called the throat), rather they tend to diverge. Hence from the perspective of the Ricci tensor this would mean $ R_{\mu \nu} n^\mu n^\nu < 0 $. The breach in attaining geodesic congruence at the throat can be manifested in terms of the violation of DEC as one can write the Ricci tensor as $ R_{\mu \nu} = T_{\mu \nu} -\frac{1}{2} T g_{\mu \nu} $. In fact, for the wormhole metric at the throat, the matter essentially violates the NEC. Rewriting the dynamic Raychaudhuri equation (\ref{eqn6}) using the transform $ K=3 \frac{\dot{\epsilon}}{\epsilon} $ one can obtain the following equation
\begin{equation}\label{eqn25}
	\ddot{\epsilon} = \frac{1}{3} (-\sigma_{\mu \nu} \sigma^{\mu \nu} - R_{\mu \nu} n^\mu n^\nu)\epsilon
\end{equation}
With $ R_{\mu \nu} n^\mu n^\nu < 0 $ there is no way of an unambiguous conclusion that the right-hand side of (\ref{eqn25}) will be negative and hence no way of ascertaining that expansions will converge. Thus essentially CNHC will not hold for such space-time. In other words, one can say that in Einstein's Gravity, regions of space-time with geometry like wormholes, that is where NEC is violated, are examples of possible regions of CNHC violation too.

\section{CNHC using the Hayward Formalism of Wormhole}\label{hayward}

The  Hayward formalism is actually the Morris Thorne wormhole, with the metric being, not necessarily asymptotically flat. The ($3+1$) dimensional wormhole metric in terms of the dual null coordinate $\xi^+$, $\xi^-$ is
\begin{equation}\label{heqn1}
	ds^2= 2g_{+-} d\xi^+ d\xi^- + r^2d\Omega^2
\end{equation}
defining the variation along the null vector as $\partial_{\pm}$ = $\frac{\partial}{\partial \xi_{\pm}}$  one can obtain the expansion as $K_{\pm} = \frac{2}{r}\partial_{\pm}r$. The notion of a trapped, untrapped and marginal sphere could be formalised based on the sign invariance of the expansion.

If the signs of $K_{+}$ and $K_{-}$  are invariant then essentially $K_{+}K_{-}>0$ and hence one will obtain a trapped sphere.

One obtains expansion concerning an untrapped sphere if $K_{+}K_{-}<0$. In this case, there are two possibilities:

\begin{itemize}
	\item $K_+ >0 ~and~ K_- < 0$, this means that the variation with respect to the null vector is outgoing for $\partial_+$ and incoming for $\partial_-$.
	\item $K_+ < 0$ and $K_->0$, this means that the variation with respect to null vector is outgoing for $\partial_-$ and incoming for $\partial_+$.
\end{itemize}

The sphere is marginal if $K_+ K_- = 0$. Then again this might be due to several possibilities listed under:
\begin{itemize}
	\item $K_+= 0$: Then $K_- < 0$ gives a future pointing marginal sphere, while $K_->0$ implies a past marginal sphere. Also, the marginal sphere is a maximal sphere if $\partial_{-}K_+ < 0$ which is classified as the outer marginal sphere. While $\partial_{-}K_+ > 0$ gives a minimal sphere that is called the inner marginal sphere.
	\item $K_+= 0, K_-= 0$: The marginal sphere is bifurcating  that is outer and inner as $\partial_{-}K_+< 0$~~or~~$\partial_{-}K_+> 0$ respectively. It is degenerate for $\partial_{-}K_+= 0$.
	\item $K_-=0$: Then $K_+< 0$ gives a future pointing marginal sphere, while $K_+> 0$ implies a past marginal sphere.
\end{itemize}

The foliation of the space-time manifold into a three-dimensional hypersurface by the marginal sphere is called the $trapping~horizon$. These trapping horizons are apparent horizons that are quasi-local causal in nature. They are classified as past/future and inner/outer just as the classification of the marginal sphere. The corresponding Einstein's equations concerning the metric (\ref{heqn1}) and energy-momentum tensor $T_{\pm}$ are given by:
\begin{equation}\label{heqn2}
	\partial_{\pm}\partial_{\pm}r-\partial_{\pm}log(-g_{+-})\partial_{\pm}r= -4\pi rT_{{\pm}{\pm}}
\end{equation}
\begin{equation}\label{heqn3}
	r\partial_+\partial_-r + \partial_+r\partial_-r -\frac{1}{2}g_{+-}= 4\pi r^2T_{{+}{-}}
\end{equation}
\begin{equation}\label{heqn4}
	r^2\partial_+\partial_- log(-g_{+-})- 2\partial_+r\partial_-r + g_{+-}= 8\pi r^2(g_{+-}T^0_0 - T_{+-})
\end{equation}

Using the definition for $ K_{\pm} $ and evaluating the first and second equations, that is equations (\ref{heqn2}) and (\ref{heqn3}), one can obtain the dynamical equations for the expansion of the null congruences as
\begin{equation}\label{heqn5}
	\partial_{\pm}K_{\pm}=-\frac{1}{2} (K_{\pm})^{2} - K_{\pm} \partial_{\pm} log(-g_{+-}) - 8\pi T_{\pm \pm}
\end{equation}
\begin{equation}\label{heqn6}
	\partial_{\pm}K_{\mp}=-K_{+}K_{-} + \frac{1}{r^{2}} g_{+-} + 8\pi T_{+-}
\end{equation}
Thus for the trapping horizon fixing $ K_{+}=0 $ on the horizon we arrive at
\begin{equation}\label{heqn7}
	\partial_{+}K_{+}= - 8\pi T_{++}
\end{equation}
\begin{equation}\label{heqn8}
	\partial_{-}K_{+}= \frac{1}{r^{2}} g_{+-} + 8\pi T_{+-}
\end{equation}
Therefore from equation (\ref{heqn8}) we can classify the future trapping horizon to be inner and outer as $ \partial_{-}K_{+}>0 $ and $ \partial_{-}K_{+}<0 $. In the case of a wormhole $ T_{+-}>0 $ and hence one would obtain a future, inner trapping horizon. The equation (\ref{heqn7}) would give us $ \partial_{+}K_{+}= - 8\pi T_{++} $. Since for wormhole we have $ T_{++}<0 $, we would end up with $ \partial_{+}K_{+}>0 $ showing that the expansion rate is positive and hence CNHC holds.

In case we fix $ K_{-}=0 $ on the horizon, we arrive at
\begin{equation}\label{heqn9}
	\partial_{-}K_{-}= - 8\pi T_{--}
\end{equation}
\begin{equation}\label{heqn10}
	\partial_{+}K_{-}= \frac{1}{r^{2}} g_{+-} + 8\pi T_{+-}
\end{equation}
Thus the trapping horizon in this case is a past outer trapping horizon with $ \partial_{-} K_{-}>0 $ since $ T_{--}<0 $. Since the expansion rate is the opposite, CNHC cannot hold here.

\section{Conclusion}
The authors of \cite{hayward2009} and \cite{moruno2009} argued that the horizon for a wormhole would be essentially ‘past outer’ type. Referring to this, it is evident that a past outer trapping horizon where CNHC is violated, allows the formation of a wormhole. Thus it verifies the results obtained in \textit{``Energy Conditions: CNHC and Wormholes"}.

Therefore, the importance of the present study is that in Einstein's Gravity it is not possible to have wormholes with normal matter but in modified gravity one may have wormholes with normal matter. Finally, we may conclude that the validity of CNHC or existence of wormhole configuration implies the non-existence of wormhole or violation of CNHC respectively but not in a reverse way both in Einstein's Gravity as well as in modified gravity theories.


\begin{thebibliography}{15}
	
	
	\bibitem{morris1988}
M. S. Morris and K. S. Thorne,
Am. J. Phys. \textbf{56}, 395 (1988)
	
	\bibitem{Lorentzian.Wormholes.Visser}
M. Visser,
Lorentzian Wormholes: From Einstein to Hawking, AIP-Press, \textbf{ISBN:978-1-56396-653-8} (1996)
	
	\bibitem{hochberg1998}
D. Hochberg and M. Visser,
Phys. Rev. Lett. \textbf{81}, 746 (1998)

	\bibitem{hochberg1998_2}
D. Hochberg and M. Visser,
Phys. Rev. D \textbf{58}, 044021 (1998)

	\bibitem{morris1988_2}
M. S. Morris, K. S. Thorne and U. Yurtsever,
Phys. Rev. Lett. \textbf{61}, 1446 (1988)

\bibitem{lobo2005}
F. S. N. Lobo,
Phys. Rev. D \textbf{71}, 084011 (2005)

\bibitem{lobo2005_2}
F. S. N. Lobo,
Phys. Rev. D \textbf{71}, 124022 (2005)

\bibitem{cataldo2012}
M. Cataldo and S. del Campo,
Phys. Rev. D \textbf{85}, 104010 (2012)

\bibitem{cataldo2013}
M. Cataldo and P. Meza,
Phys. Rev. D \textbf{87}, 064012 (2013)

\bibitem{folomeev2014}
V. Folomeev and V. Dzhunushaliev,
Phys. Rev. D \textbf{89}, 064002 (2014)

\bibitem{ali1}
G. Mustafa, M. Ahmad, A. \"Ovg\"un, M. F. Shamir and I. Hussain, Fortschr. Phys. \textbf{69}, 2100048 (2021)

\bibitem{ali2}
A. \"Ovg\"un, K. Jusufi and \.I. Sakallı, Phys. Rev. D \textbf{99}, 024042 (2019)

\bibitem{ali3}
W. Javed, R. Babar and A. \"Ovg\"un, Phys. Rev. D \textbf{99}, 084012 (2019)

\bibitem{ali4}
M. G. Richarte, I. G. Salako, J. P. Morais Graca, H. Moradpour and A. \"Ovg\"un, Phys. Rev. D \textbf{96}, 084022 (2017)

\bibitem{ali5}
A. \"Ovg\"un, Phys. Rev. D \textbf{98}, 044033 (2018)

\bibitem{ali6}
A. \"Ovg\"un, Eur. Phys. J. Plus \textbf{136}, 987 (2021)

\bibitem{ali7}
M. Halilsoya, A. \"Ovg\"un, S. H. Mazharimousavi, Eur. Phys. J. C \textbf{74}, 2796 (2014)

%\bibitem{ali8}
%A. \"Ovg\"un, Turk. J. Phys. \textbf{44: 5}, 465 (2020)

\bibitem{moraes}
P. H. R. S. Moraes and P. K. Sahoo, Phys. Rev. D \textbf{96}, 044038 (2017)

\bibitem{moraes2}
G. Mustafa, Z. Hassan, P. H. R. S. Moraes and P. K. Sahoo, Phys. Lett. B \textbf{821}, 136612 (2021)

\bibitem{matter1}
S. D. Forghani and S. H. Mazharimousavi, J. Cos. Astro. Phys. \textbf{11}(2020)018 (2020)

\bibitem{matter2}
A. Ditta, I. Hussain, G. Mustafa, A. Errehymy and M. Daoud, Eur. Phys. J. C \textbf{81}, 880 (2021)

\bibitem{matter3}
R. Sengupta, S. Ghosh, M. Kalam and S. Ray, Class. Quant. Grav. \textbf{39}, 105004 (2022)

\bibitem{matter4}
R. Sengupta, S. Ghosh, M. Kalam and S. Ray, Int. J. Geom. Meth. Mod. Phys. \textbf{19: 02}, 2250019 (2022)

	\bibitem{gibbons1977}
G. W. Gibbons and S. W. Hawking,
Phys. Rev. D \textbf{15}, 2738 (1977)

	\bibitem{hawking1982}
S. W. Hawking and I. L. Moss,
Phys. Lett. B \textbf{110: 1}, 35-38 (1982)

	\bibitem{wald1983}
R. M. Wald,
Phys. Rev. D \textbf{28}, 2118 (1983)

	\bibitem{kitada1992}
Y. Kitada and K. Maeda,
Phys. Rev. D \textbf{45}, 1416 (1992)

	\bibitem{chakraborty2001}
S. Chakraborty and B. C. Paul,
Phys. Rev. D \textbf{64}, 127502 (2001)

	\bibitem{chakraborty2002}
S. Chakraborty and S. Chakrabarti,
Class. Quant. Grav. \textbf{19}, 3775 (2002)

	\bibitem{chakraborty2003}
S. Chakraborty and U. Debnath,
Class. Quant. Grav. \textbf{20}, 2693 (2003)

	\bibitem{chakraborty2009}
S. Chakraborty and T. Bandyopadhyay,
Gen. Relt. Grav. \textbf{41}, 2461–2467 (2009)

	\bibitem{hayward1998}
S. A. Hayward,
Class. Quant. Grav. \textbf{15}, 3147 (1998)

	\bibitem{hayward2004}
S. A. Hayward and H. Koyama,
Phys. Rev. D \textbf{70}, 101502 (2004)

	\bibitem{hayward2009}
S. A. Hayward,
Phys. Rev. D \textbf{79}, 124001 (2009)

	\bibitem{moruno2009}
P. Martin-Moruno and P. F. Gonzalez-Diaz,
Phys. Rev. D \textbf{80}, 024007 (2009)








	
\end{thebibliography}
\end{document}